\documentstyle{laa}

\newcommand{\rmag}    {${\rlap.}^{m}$}
\newcommand{\lsi}    {LSI+61$^{\circ}$303}

\newcommand{\kms}    {km~s$^{-1}$}
\newcommand{\ltsima} {$\; \buildrel < \over \sim \;$}
\newcommand{\simlt}  {\lower.5ex\hbox{\ltsima}}            
\newcommand{\gtsima} {$\; \buildrel > \over \sim \;$}
\newcommand{\simgt}  {\lower.5ex\hbox{\gtsima}}            

\begin{document}

\thesaurus{06(08.09.2;08.05.2;08.22.3;13.18.5;13.25.5)}

\title{Photometric and H$\alpha$ observations of \lsi:\\
detection of a $\sim$26 day V and JHK band modulation}

\author{J.~M.~Paredes\inst{1,2}
\and P.~Marziani\inst{3}
\and J.~Mart\'{\i}\inst{1}
\and J.~Fabregat\inst{4}
\and M.~J. Coe\inst{5}
\and C. Everall\inst{5}
\and F. Figueras\inst{1,2}
\and C. Jordi\inst{1,2}
\and A.~J. Norton\inst{6}
\and T. Prince\inst{7}
\and V. Reglero\inst{4}
\and P. Roche\inst{5}
\and J. Torra\inst{1,2}
\and S.~J. Unger\inst{6}
\and R. Zamanov\inst{8}
}
\institute{Departament d'Astronomia i Meteorologia, Universitat de
Barcelona, Av.  Diagonal 647, E-08028 Barcelona, Spain
\and
And Laboratori d'Astrof\'{\i}sica, Societat Catalana de F\'{\i}sica (IEC),
Spain
\and
Department of Physics \& Astronomy, University of Alabama,
Tuscalosa AL 35487-0324, USA
\and
Departamento de Matem\'atica Aplicada y Astronom\'{\i}a, Universidad de
Valencia,
46100 Burjassot, Valencia, Spain
\and
Physics Department, University of Southampton,
Southampton SO9 5NH, UK
\and
Department of Physics, The Open University,
Walton Hall, Milton Keynes MK7 6AA, UK
\and
Division of Physics, Mathematics and Astronomy, Caltech, Pasadena, CA 91125,
USA
\and
National Astronomical Observatory Rozhen,
POB 136, 4700 Smoljan, Bulgaria}

\offprints{J.~M.~Paredes}

\date{Received           ; Accepted                  }

\maketitle

\begin{abstract}

We present new optical and infrared photometric observations and high
resolution H$\alpha$ spectra of the periodic radio star \lsi. The
optical photometric data set covers the time interval 1985-1993 and
amounts to about a hundred nights.  A period of $\sim$26 days is found
in the V band.  The infrared data also present evidence for a similar
periodicity, but with higher amplitude
of variation (0\rmag 2). The spectroscopic
observations include 16 intermediate and high dispersion spectra of
\lsi\ collected between January 1989 and February
1993.  The H$\alpha$ emission line profile and its variations are
analyzed.  Several emission line parameters -- among them the
H$\alpha$ EW and the width of the H$\alpha$ red hump -- change
strongly at or close to radio maximum, and may exhibit periodic
variability.  We also observe a significant change in the peak
separation.  The H$\alpha$ profile of \lsi\ does not seem peculiar
for a Be star.  However, several of the observed variations of the
H$\alpha$ profile can probably be associated with the presence of the
compact, secondary star.

\end{abstract}


\section{Introduction} \label{intro}

The early-type star \lsi\ (V~615~Cas) is the optical counterpart of the
variable radio source GT~0236+610, discovered during a galactic plane
radio survey (Gregory \& Taylor, 1978).  Taylor \& Gregory (1982) found
that this object exhibits strong radio outbursts with a 26.5~d period.
Further observations (Taylor \& Gregory, 1984) established the
currently accepted value of 26.496$\pm$0.008~d.  Typically, radio
outbursts peak around phases 0.6-0.8 (Paredes et al., 1990).  The
spectroscopic radial velocity observations of Hutchings \& Crampton
(1981), hereafter HC81, are in agreement with the radio period, and give
support to the presence of a companion.  In addition, they also conclude
that the optical spectrum corresponds to a rapidly rotating B0 V star,
with an equatorial disk and mass loss.

All the radio data available to date on the outburst peak flux density
provide evidence for a strong modulation, over a time scale of 4~yr, in
the amplitude of the 26.5~d periodic radio outbursts (Gregory et al.,
1989; Paredes et al., 1990; Estalella et al., 1993).  The dependence of
radio outbursts flux density on frequency, the peak time delay, and the
general shape of the radio light curves can be modeled as continuous
relativistic particle injection into an adiabatically expanding
synchrotron emitting source (Paredes et al., 1991).  Furthermore, recent
VLBI observations have provided the first high resolution map of
\lsi\, showing a double sub-arcsec structure (Massi et al.,
1993).  The physical parameters derived from these VLBI observations
and those of Taylor et al. (1992) are in agreement with this model.

The system was detected as an X-ray source by Bignami et al. (1981)
and has also been proposed (Gregory \& Taylor, 1978; Perotti et al., 1980)
to be the radio counterpart of the COS B $\gamma$-ray source
CG135+01 (Hersem et al., 1977). However, this last association is still
doubtful due to the large $\gamma$-ray error box.

Paredes \& Figueras (1986) based on
UBVRI photometric observations detected
optical variability roughly correlated with the radio light curve.  The
amplitude was about 0\rmag 1. A model based on deformations of the
primary star by a compact companion in a eccentric system was initially
applied by Paredes (1987) to explain it.  Optical variability with time
scales of days has also been reported by Lipunova (1988) who, in
addition, found short term nightly fluctuations of some hundredths of a
magnitude.  These short time fluctuations were first observed by
Bartolini et al.  (1983).  More recently, Mendelson \& Mazeh (1989)
reported an optical modulation with amplitude similar to that found by
Paredes \& Figueras (1986) and with a period of 26.62~$\pm$~0.09~d, near
to the radio value, in the Johnson I band.  However, their data set was
not sufficient to show clearly a similar periodicity at shorter
wavelengths.

The photometric results presented in this paper confirm that a $\sim$26~d
periodicity is actually present in the V band. In addition,
the general shape of the visual light curve is very similar to that
observed by Mendelson \& Mazeh (1989) in the I band.
In the JHK near infrared bands, we find clear evidence that a similar periodic
modulation is also present, with amplitude of $\sim$0\rmag 2.

On the other hand, intermediate resolution spectroscopic data of \lsi\
suitable for an analysis of the H$\alpha$ emission line profile are
available in literature, at the time of writing, only from the early
papers of Gregory et al.  (1979) and HC81.  These papers outlined the
variations of the H$\alpha$ profile, but were far from reaching any
firm conclusion regarding the mechanism responsible for such variations.
In this work, we present 16 spectra (9 of them of high resolution,
0.2-0.44 \AA\ FWHM) obtained with linear detectors.  The new spectra
allow a more detailed description of the H$\alpha$ line profile and
of its variation.  We confirm many of the early findings of Gregory et
al.  (1979) and of HC81 concerning line shifts and H$\alpha$ EW
variations.  Moreover, we find that other intriguing changes in the
H$\alpha$ line profile (noticeably the width of the red hump) occur
at or close to radio outburst.

\section{Photometric observations and results} \label{phot}

The Johnson photometric observations were made at Calar Alto (Almer\'{\i}a,
Spain) with the 1.23~m telescope of the Centro Astron\'omico Hispano-Alem\'an
(CAHA) and the 1.52~m telescope of the Observatorio Astron\'omico
Nacional (OAN) and at the Observatorio del Roque de los Muchachos
(ORM, La Palma, Spain), using the 1~m Jacobus Kapteyn telescope (JKT).
They cover the period 1985-1993 and amount to one hundred independent
photometric measurements.  Both Calar Alto telescopes are equipped with
a one channel photometer with a dry-ice cooled RCA 31034
photomultiplier.  The JKT observations were made
using the People's photometer, with two channels, which is equipped with
EMI~9658AM photomultipliers.

The differential photometry was performed using SAO 12319 (V=8.79 and
I=7.76), SAO 12327 (V=8.15 and I=6.90) and BD+60$^{\circ}$493 (V=8.41
and I=7.04) as comparison stars.  The majority of measurements were
obtained in the Johnson V filter, although some simultaneous I filter
observations were also taken and will be reported here.  Differences of
magnitude between comparison stars themselves are constant within 0\rmag 02.
Further details of the observing technique are reported in Paredes
\& Figueras (1986).

The infrared observations were made at the Teide Observatory, (Tenerife,
Spain),
using the 1.5~m Carlos S\'anchez telescope (TCS) equipped with the continuously
variable filter (CVF) photometer. The data were corrected for atmospheric
extinction and flux-calibrated by comparison with an adequate sample of
standard stars.

\begin{figure}
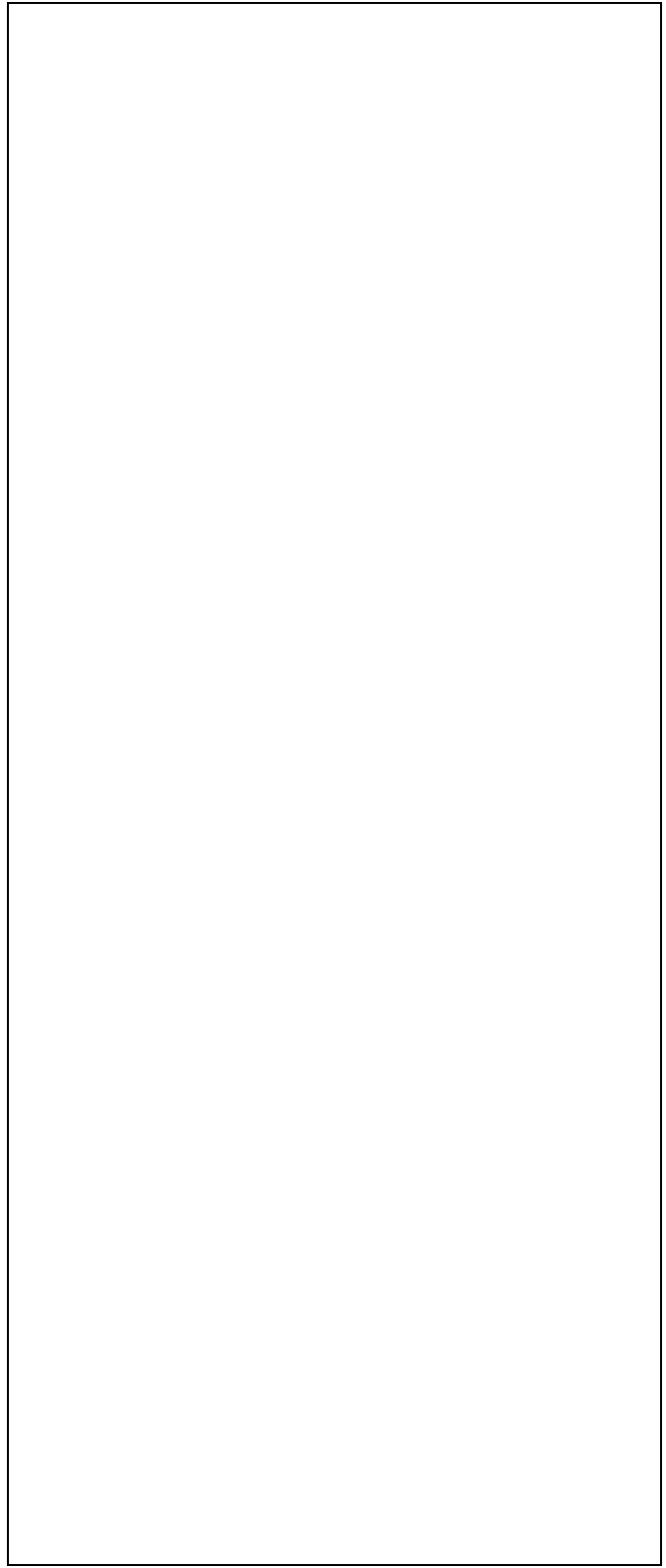

\picplace{21.0cm}
\caption{\label{mos}
Optical and infrared observations of \lsi\ folded on the 26.496~d radio period.
Phase zero has been set at JD 2443366.775 (Taylor \& Gregory, 1982).
{}From top to bottom, the V, I, J, H, and K band are
plotted. The dots at phase 0.7 are from Elias et al. (1985).
All observations are plotted twice.}
\end{figure}

The results of our Johnson photometric observations are given in Table
\ref{vi}.
First column indicates the Julian date,
second and third column are,
respectively, the Johnson V and I band magnitudes of \lsi.
A similar format has been used for the infrared observations,
whose results are presented in Table \ref{ir}.
Figure \ref{mos} shows, with the same scale, the optical and infrared
full data set folded with the radio period of 26.496~d.

\begin{table}
\caption[ ]{\label{vi} Johnson photometric observations of \lsi}
\begin{tabular}{cccccc}
\hline
                   &          &          &                   &          &
    \\
Julian Date        & V        & I        &  Julian Day       & V        & I
    \\
(2440000+)         &          &          &  (2440000+)       &          &
    \\
                   &          &          &                   &          &
    \\
\hline
6264.62            & 10.74    &  9.26    &  7452.68          & 10.76    &   -
    \\
6265.59            & 10.74    &  9.25    &  7456.44          & 10.77    &   -
    \\
6267.55            & 10.75    &  9.24    &  7456.63          & 10.77    &   -
    \\
6268.58            & 10.70    &  9.20    &  7802.60          & 10.77    &  9.31
    \\
6269.61            & 10.72    &  9.20    &  7803.61          & 10.78    &  9.32
    \\
6270.62            & 10.65    &  9.14    &  7804.60          & 10.79    &  9.31
    \\
6271.60            & 10.71    &  9.20    &  7805.59          & 10.77    &  9.29
    \\
6272.60            & 10.70    &  9.19    &  7806.64          & 10.77    &  9.29
    \\
6273.61            & 10.74    &  9.23    &  7807.64          & 10.77    &  9.28
    \\
6274.60            & 10.74    &  9.24    &  7808.62          & 10.77    &  9.29
    \\
6421.52            & 10.77    &  9.30    &  7809.59          & 10.76    &  9.27
    \\
6423.34            & 10.73    &  9.28    &  7833.67          & 10.75    &  9.28
    \\
7056.45            & 10.76    &  9.27    &  7837.59          & 10.74    &  9.29
    \\
7058.43            & 10.74    &  9.24    &  7838.56          & 10.72    &  9.28
    \\
7060.58            & 10.76    &  9.27    &  7957.34          & 10.74    &  9.32
    \\
7063.57            & 10.73    &  9.24    &  8176.43          & 10.75    &  9.30
    \\
7069.64            & 10.72    &  9.21    &  8180.38          & 10.74    &  9.28
    \\
7096.66            & 10.64    &   -      &  8238.30          & 10.76    &  9.31
    \\
7097.56            & 10.65    &   -      &  8274.46          & 10.79    &   -
    \\
7098.64            & 10.70    &   -      &  8275.49          & 10.79    &   -
    \\
7125.35            & 10.68    &   -      &  8474.69          & 10.79    &  9.28
    \\
7128.30            & 10.67    &   -      &  8527.46          & 10.75    &  9.29
    \\
7147.30            & 10.75    &   -      &  8545.53          & 10.78    &   -
    \\
7148.30            & 10.76    &   -      &  8620.50          & 10.75    &   -
    \\
7149.59            & 10.71    &   -      &  8898.65          & 10.76    &  9.29
    \\
7151.40            & 10.73    &   -      &  9003.46          & 10.74    &  9.28
    \\
7152.43            & 10.72    &   -      &  9004.47          & 10.75    &  9.29
    \\
7169.28            & 10.74    &   -      &  9005.45          & 10.75    &  9.28
    \\
7170.45            & 10.70    &   -      &  7125.35          & 10.67    &   -
    \\
7169.54            & 10.71    &   -      &  7128.30          & 10.68    &   -
    \\
7171.56            & 10.67    &   -      &  7168.43          & 10.80    &   -
    \\
7172.35            & 10.71    &   -      &  7169.30          & 10.78    &   -
    \\
7172.57            & 10.64    &   -      &  7170.44          & 10.78    &   -
    \\
7231.34            & 10.74    &   -      &  7367.64          & 10.75    &   -
    \\
7232.35            & 10.70    &   -      &  7368.64          & 10.80    &   -
    \\
7381.65            & 10.74    &   -      &  7531.32          & 10.73    &   -
    \\
7381.65            & 10.74    &   -      &  8092.58          & 10.77    &   -
    \\
7388.63            & 10.79    &   -      &  8194.49          & 10.71    &   -
    \\
7389.59            & 10.78    &   -      &  8207.51          & 10.76    &   -
    \\
7390.59            & 10.77    &   -      &  8208.42          & 10.74    &   -
    \\
7391.64            & 10.77    &   -      &  8209.35          & 10.70    &   -
    \\
7392.67            & 10.78    &   -      &  8262.55          & 10.75    & 9.31
    \\
7393.59            & 10.78    &   -      &  8266.31          & 10.73    & 9.31
    \\
7394.59            & 10.75    &   -      &  8267.31          & 10.74    & 9.28
    \\
7395.65            & 10.76    &   -      &  8268.30          & 10.74    & 9.33
    \\
7420.54            & 10.70    &   -      &  8270.31          & 10.76    & 9.34
    \\
7421.52            & 10.74    &   -      &  8577.53          & 10.79    &   -
    \\
7422.56            & 10.74    &   -      &  8610.38          & 10.76    &   -
    \\
7422.69            & 10.74    &   -      &  8611.41          & 10.76    &   -
    \\
7423.65            & 10.77    &   -      &  8962.45          & 10.79    &   -
    \\
7424.64            & 10.79    &   -      &  9034.40          & 10.74    &   -
    \\
7425.59            & 10.79    &   -      &  9034.40          & 10.69    &   -
    \\
7426.62            & 10.78    &   -      &                   &          &
    \\
\hline
\end{tabular}
\end{table}

\section{Photometric analysis} \label{photan}

\subsection{V Photometric periodicity}  \label{vper}

In order to try to confirm independently the periodic optical modulation
reported by Mendelson \& Mazeh (1989), a period analysis was carried out
over this entire data set, amounting to 105 photometric V measurements
over the time interval 1985-1993.  The period analysis of the data was
performed by using the phase dispersion minimization (PDM) technique
(Stellingwerf, 1978).  This method consists of assuming a trial period
and then constructing a phase diagram.  The phase interval is divided
into bins and the variance of the data points is computed in each bin.
The weighted mean of the variances is divided by the total variance of
the data.  It can be shown that local minima of this function correspond
to periods present in the data or to multiples of such periods.

\begin{figure}
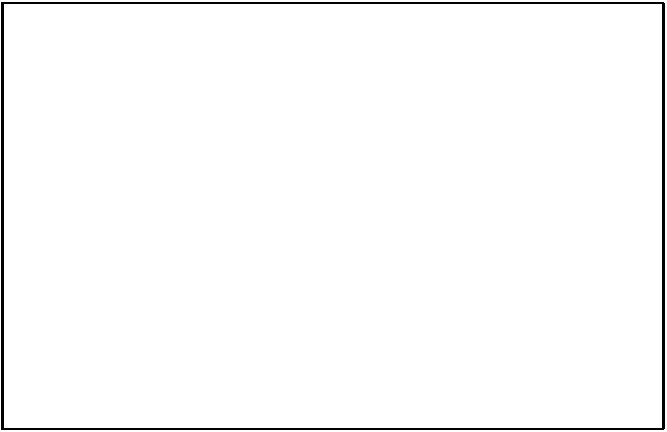

\picplace{5.65cm}
\caption{\label{pdmv} PDM periodogram of the V band data with
bin structure (5,5) and frequency step
$5\times 10^{-5}$~c~d$^{-1}$. The frequencies are in cycles per day.
The deepest minimum occurs at a frequency of 0.0387~c~d$^{-1}$,
corresponding to a period of 25.8~d.}
\end{figure}

Considering that our minimum sampling rate is about one day, our period
search was made up to a frequency of 0.5 c~d$^{-1}$.
The result of PDM analysis of the V band data is shown in Fig. \ref{pdmv}.
The most significant minimum in the explored frequency range (from 0.01
to 0.5 c~d$^{-1}$) occurs on 0.0387~c~d$^{-1}$ and the uncertainty that
we associate with this frequency is the frequency resolution of the
complete data set, given by $\sim1/T$, where $T$ is the total length of
the data span and is equal to 0.0004~c~d$^{-1}$.  This corresponds to a
period of 25.8$\pm$0.3 d.

\begin{figure}
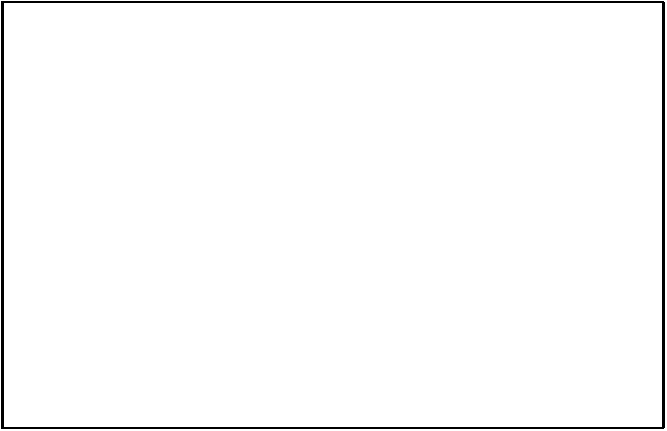

\picplace{5.65cm}
\caption{\label{vmig} Bin average V light curve of \lsi\
folded on the 26.496~d radio period.
Phase zero has been set at JD 2443366.775 (Taylor \& Gregory, 1982).
Errors bars indicate the formal
estimate of the uncertainty of the mean within each bin. The continuous
line is plotted for visual aid. All data is plotted twice.}
\end{figure}

{}From this analysis, it appears evident that a
modulation with period of $\sim$26~d is actually present in the Johnson
V photometry of \lsi.  This period is similar
to that of 26.62~d found by Mendelson \& Mazeh (1989) in the Johnson I band
and to the 26.496~d radio periodicity (Taylor \& Gregory, 1984).  In
Fig.~\ref{vmig} we have plotted our 105 photometric V points folded on the
26.496~d radio period and binned into 10 bins. Error
bars indicate the formal estimate of the uncertainty of the mean within
each bin. For clarity, the data are plotted twice.
{}From this figure we note the similar shape with the optical light curves
presented by Mendelson \& Mazeh (1989). In particular, the presence
of a broad brightness maximun near radio active phases 0.5-0.9, and
a clear minimum around phase 0.3 can be appreciated.

Once the existence of an optical modulation with a period near to
26~d has been established in an independent way from Mendelson \& Mazeh
(1989), it is worth carrying out an analysis of all long term
photometric data available today for \lsi.  In this way, we have
searched for periodicities, in the range from 20 to 30~d, the ensemble
consisting of both our data and the photometric points published by
Bartolini et al.  (1983), Lipunova (1988) and Mendelson \& Mazeh (1989),
amounting to 204 nights.  Using a frequency step of $2\times
10^{-7}$~c~d$^{-1}$ and a bin structure (5,2), the deepest PDM minimum
corresponds to 26.5$\pm$0.2~d, although it is not very prominent.
This period value is coincident with the radio period.

\subsection{I Photometry}  \label{iphot}

Our set of Johnson I band observations, listed in Table \ref{vi}, amount to
43 nights only.  With this small data set, it is not possible to carry
out a feasible periodicity search.  In Fig.~\ref{mos}, we show the I band
observations folded with the 26.496~d radio period.  The available data
covers the phase interval 0.2-0.9~.  However, this partial light curve
presents the same trends as that of V band observations.  In
particular, a maximum near the central radio phases and low emission
level near phase 0.2 is clearly seen.

\subsection{JHK bands photometric periodicity} \label{irper}

Our JHK photometric observations are plotted in Fig.~\ref{mos} as a function of
radio phase.  Also, we have included two points (dots) at phase
$\sim0.7$ observed by Elias et al.  (1985).  The infrared data
available have a good coverage over the full radio period, and indicates
that the infrared light curves of \lsi\ also present a modulation
similar to that of V and I bands.  However, the infrared high emission
state (orbital phases $\sim$0.6-0.9) is broader than in the optical,
while the minimum emission state (orbital phases $\sim$0.2-0.4) is
deeper and narrower.

\begin{table}
\begin{center}
\caption[ ]{\label{ir} JHKL' photometric observations of \lsi}
\begin{tabular}{ccccc}
\hline
                &          &          &           &           \\
Julian Day      & J        & H        &  K        & L'        \\
(2440000+)      &          &          &           &           \\
                &          &          &           &           \\
\hline
  7161.46   & 8.76     & 8.39     & 8.06      &  -        \\
  7163.45   & 8.81     & 8.39     & 8.13      & 7.60      \\
  7164.38   & 8.77     & 8.38     & 8.15      & 7.59      \\
  8279.45   & 8.73     & 8.32     & 8.03      &  -        \\
  9013.47   & 8.62     & 8.25     & 7.88      &  -        \\
  8284.51   & 8.60     & 8.30     & 7.90      &  -        \\
  8491.59   & 8.74     & 8.34     & 8.04      &  -        \\
  8493.57   & 8.80     & 8.35     & 8.05      &  -        \\
  8494.60   & 8.68     & 8.26     & 7.97      &  -        \\
  8496.55   & 8.62     & 8.22     & 7.91      &  -        \\
  8498.51   & 8.57     & 8.18     & 7.87      &  -        \\
  8590.35   & 8.70     & 8.20     & 7.90      &  -        \\
  8592.47   & 8.71     & 8.29     & 8.01      &  -        \\
  8665.42   & 8.57     & 8.16     & 7.89      &  -        \\
  8669.46   & 8.60     & 8.25     & 7.96      &  -        \\
  8856.65   & 8.66     & 8.20     & 7.88      &  -        \\
  8857.67   & 8.62     & 8.21     & 7.92      &  -        \\
  8858.65   & 8.62     & 8.19     & 7.92      &  -        \\
\hline
\end{tabular}
\end{center}
\end{table}

\begin{figure}
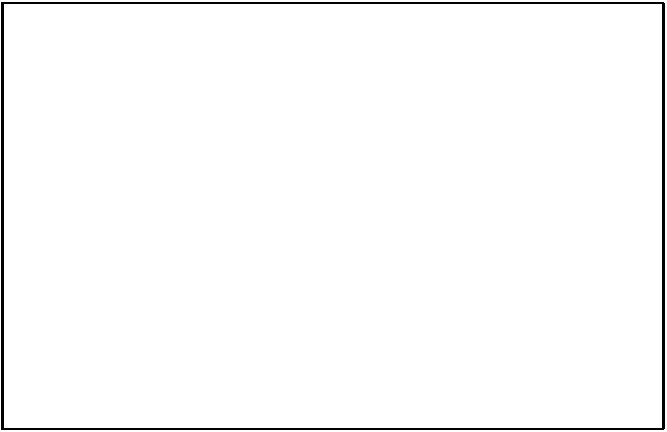

\picplace{5.65cm}
\caption{\label{PDMJHK} PDM periodogram of the merged JHK band data with
bin structure (5,5) and frequency step
$5\times 10^{-5}$~c~d$^{-1}$. The frequencies are in cycles per day.
The deepest minimum occurs at a frequency of 0.0370~c~d$^{-1}$,
corresponding to a period of 27.0~d.}
\end{figure}

\begin{figure}
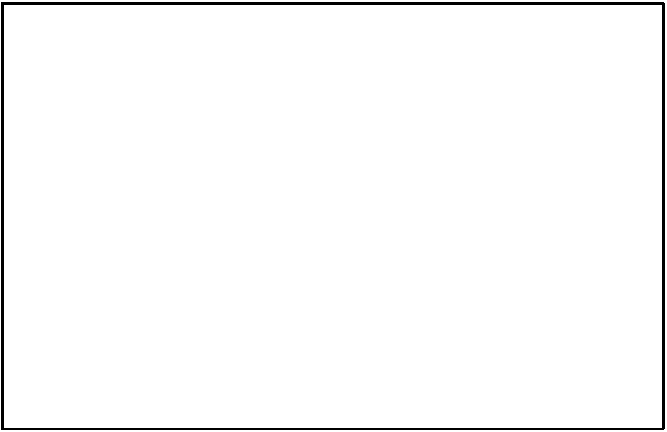

\picplace{5.65cm}
\caption{\label{jhkmer} JHK infrared observations of \lsi\ after substracting
the mean
magnitude of each filter and normalizing with their respective r.m.s.
dispersion. They are shown folded on the 26.496~d radio period.
Phase zero  has been set at JD 2443366.775.
All observations are plotted twice. }
\end{figure}

For a single infrared band, the amount of data accumulated by us, 18
nights, could not suffice to carry out a significant period search.  In
order to overcome this problem, we have merged all the JHK photometric
points after subtracting their respective mean and dividing by the
corresponding r.m.s. dispersion in each filter.  This process is roughly
equivalent to having a very broad bandpass filter, about 1$\mu$m wide.
The two points observed by Elias et al.  (1985) have also been included.
This provides a data set of relative normalized infrared magnitudes,
with 60 measurements over which we have carried a PDM period analysis.
The minimum sampling rate is about one day.  So, the PDM search was
carried out up to a frequency of 0.5 c~d$^{-1}$.  The result of PDM
analysis of the merged JHK band data is shown in Fig.  \ref{PDMJHK}.
The most prominent minimum occurs on $0.0370\pm0.0003$ c d$^{-1}$,
corresponding to a period of $27.0 \pm 0.3$ d. Another nearby deep minimum,
with comparable significance, is found at $0.0376\pm0.0003$ c d$^{-1}$,
corresponding to a period of $26.6 \pm 0.3$ d.
This implies that an
infrared modulation, with period similar to the radio period, is
also present in \lsi.
The 60 normalized
points used are plotted in Fig.  \ref{jhkmer} as a function of radio
phase, computed using the radio period value of 26.496 d.

\subsection{Photometric discussion} \label{photdis}

The JHK infrared light curves of Fig. \ref{mos}, showing a deep minimum at
phase $\sim$0.3 and a rather flat maximum centered around phase $\sim$0.8,
are reminiscent of light curves from eclipsing variables.  From a
rotation velocity value of $v\sin{i}\simeq360\pm25$ km~s$^{-1}$, HC81
suggest that orbital inclination of \lsi\ is close to
$90^{\circ}$.  So, this makes the eclipse possibility a rather
reasonable interpretation.

In addition, the~existence~of~an IR excess in
LSI+61$^{\circ}$303
has been reported by
D'Amico et al.\  (1987) and Elias et al.\  (1985).  This is, however, a
rather common situation in Be stars, where the IR excess at micron
wavelengths is attributed to a dense circumstellar envelope (Slettebak,
1979).  This envelope can be also partially responsible for the
H$\alpha$ emission.  Due to the presence of this envelope, its
free-free and free-bound opacity is also very likely to absorbe the
infrared radiation from any orbiting companion, thus strongly influencing
the observed light curve beyond a simple geometrical eclipse.
A theoretical modelling of the JHK light curves, based on this
eclipse-attenuation scenario, could yield a determination of the system
orbital parameters (Mart\'{\i} \& Paredes, 1994).

On the other hand, when considering the optical light curves, one sees
that the minimum is wider and lasts for about half an orbital cycle (see
Fig. \ref{vmig}).  Then, as suggested by Mendelson \& Mazeh (1989), the eclipse
explanation at these wavelengths is more difficult to accept.  However,
due to the approximate frequency dependence $\tau_{\nu} \propto
\nu^{-3}$, the optical free-free and free-bound extinction is actually
very low.  So, any attenuation effects will be difficult to appreciate
in the optical band, meaning that optical variability should be
accounted for by a different physical mechanism, probably involving
X-ray heating of parts of the normal star facing the X-ray source.

\section{Spectroscopic observations and results} \label{spec}

Sixteen spectra covering the H$\alpha$ spectral range were collected
during several observing runs from January 1989 to February 1993.  Table
\ref{speclog} reports the dates of observation and basic information on
the instrumental setup.  The first column indicates the spectrum
identification, the second and third the observatory and telescope used,
the fourth, fifth and sixth contain the date, UT time and Julian day of
observation, respectively.  Seventh column is the radio phase.  Finally,
the eigth and ninth columns indicate the dispersion and covered spectral
range.  All spectra were recorded employing CCD detectors.  They were
bias subtracted and flat field corrected using the IRAF or FIGARO
package, with the exception of Rozhen spectra which have been reduced
using the pcIPS software package (Smirnov et al., 1992).  Only the
spectrum obtained on Dec. 27, 1990 (LP1, see Table \ref{speclog}) at La
Palma was flux calibrated.

\begin{table*}
\caption[ ]{\label{speclog} Summary of spectroscopic observations of \lsi}
\begin{tabular}{clccccccc}
\hline
Id. & Observatory & Telescope & Date        &  UT            & Julian Day  &
Radio & Dispersion & Spectral Range \\
    &             &           &             &                & (2440000+)  &
Phase & (\AA/pix)  & (\AA)          \\
\hline
AS1 & Asiago      &    1.8m   & 1989 Jan 18 &  19$^h$29$^m$  &   7545.3    &
0.70  &    0.22    & 6513-6650      \\
AS2 & Asiago      &    1.8m   & 1989 Jan 19 &  19$^h$27$^m$  &   7546.3    &
0.74  &    0.22    & 6510-6652      \\
AS3 & Asiago      &    1.8m   & 1989 Jan 19 &  20$^h$21$^m$  &   7546.3    &
0.74  &    0.22    & 6510-6652      \\
AS4 & Asiago      &    1.8m   & 1989 Jan 21 &  19$^h$55$^m$  &   7548.3    &
0.82  &    0.22    & 6510-6652      \\
AS5 & Asiago      &    1.8m   & 1989 Jan 21 &  20$^h$47$^m$  &   7548.4    &
0.82  &    0.22    & 6510-6652      \\
LP1 & La Palma    &    2.5m   & 1990 Dec 27 &  01$^h$14$^m$  &   8252.5    &
0.40  &    0.38    & 6492-6672      \\
LP2 & La Palma    &    2.5m   & 1991 Jan 27 &  23$^h$18$^m$  &   8283.5    &
0.57  &    0.78    & 6400-6800      \\
LP3 & La Palma    &    2.5m   & 1991 Aug 28 &  02$^h$05$^m$  &   8497.5    &
0.64  &    0.36    & 6472-6680      \\
MP1 & Mt. Palomar &    1.5m   & 1992 Aug 17 &  08$^h$05$^m$  &   8851.8    &
0.02  &    1.00    & 6255-6925      \\
MP2 & Mt. Palomar &    1.5m   & 1992 Aug 17 &  08$^h$13$^m$  &   8851.8    &
0.02  &    1.00    & 6255-6925      \\
MP3 & Mt. Palomar &    1.5m   & 1992 Aug 18 &  09$^h$57$^m$  &   8852.9    &
0.06  &    1.00    & 6285-6925      \\
MP4 & Mt. Palomar &    1.5m   & 1992 Aug 19 &  10$^h$04$^m$  &   8853.9    &
0.09  &    1.00    & 6255-6935      \\
RO1 & Rozhen      &    2.0m   & 1992 Sep 03 &  23$^h$30$^m$  &   8869.5    &
0.68  &    0.10    & 6530-6589      \\
RO2 & Rozhen      &    2.0m   & 1992 Oct 09 &  21$^h$15$^m$  &   8905.4    &
0.04  &    0.10    & 6543-6602      \\
RO3 & Rozhen      &    2.0m   & 1993 Feb 05 &  17$^h$40$^m$  &   9024.2    &
0.52  &    0.10    & 6533-6594      \\
RO4 & Rozhen      &    2.0m   & 1993 Feb 06 &  19$^h$30$^m$  &   9025.3    &
0.56  &    0.10    & 6534-6594      \\
\hline
\end{tabular}
\end{table*}

\begin{figure}
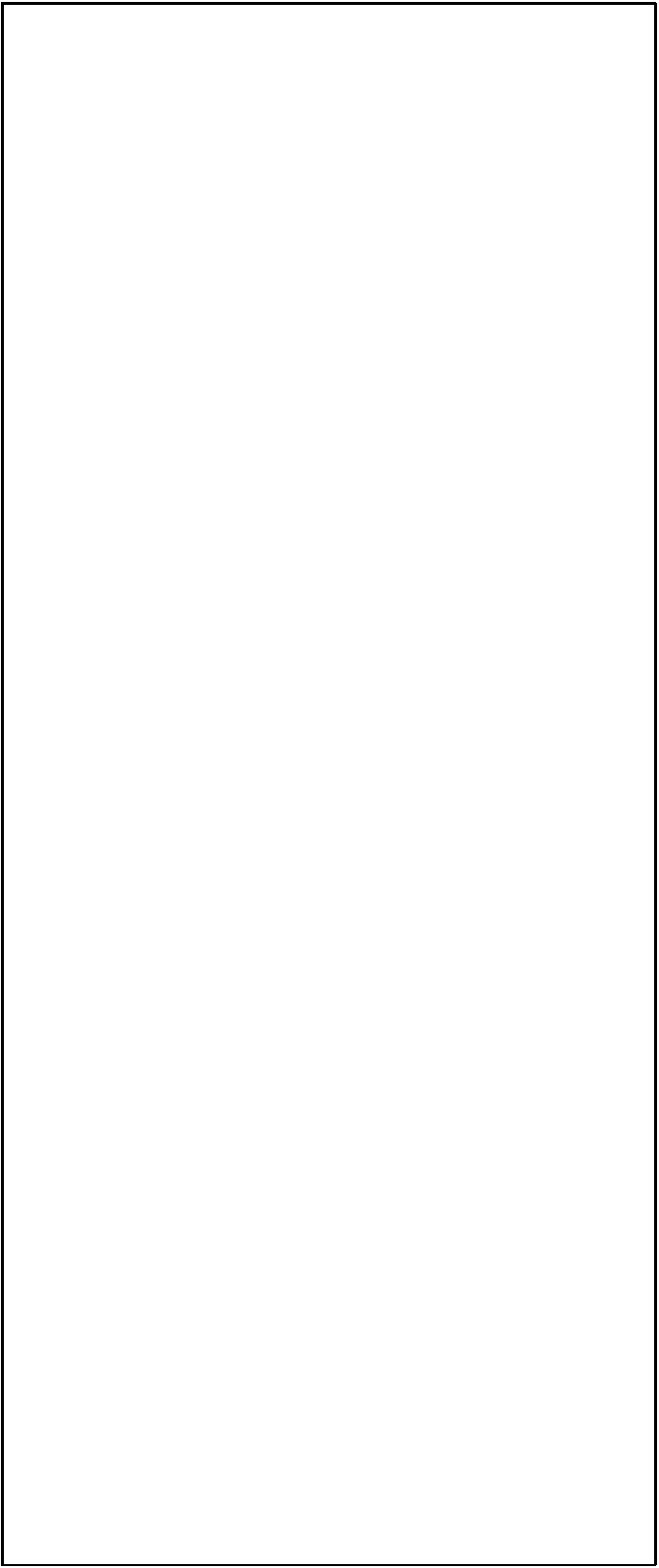

\picplace{21.00cm}
\caption{\label{specplot} Normalized H$\alpha$ line profiles of \lsi\ at
different
radio phases given on right-hand side.
Horizontal scale
is wavelength in \AA~and vertical scale is in arbitrary intensity units.
The Rozhen Observatory spectra
have been convolved with a gaussian of FWHM = 0.3 \AA.
Identification codes are revealed in Table 3.}
\end{figure}

\begin{table*}
\caption[ ]{\label{specpar} Observed H$\alpha$ line parameters of \lsi}
\begin{tabular}{cccccccccc}
\hline
Id. & $v_r$(B) &$v_r$(dip)& $v_r$(R) & FWHM(B) & FWHM(R) & B/R   &
$-$EW(H$\alpha$)    & EW(wings) & EW(R)/EW(B) \\
    & (\kms  ) & (\kms)   & (\kms)   & (\AA)   &(\AA)    &       &   (\AA)
       &  (\AA)    &             \\
\hline
AS1 & $-$227   & $-$50    & 154      &   5.10  &  6.85   & 1.14  &       11.46
       &  1.96     & 0.79        \\
AS2 & $-$208   & $-$62    & 130      &   5.57  &  8.67   & 1.18  &        6.26
       &  0.61     & 0.73        \\
AS3 & $-$228   & $-$56    & 125      &   5.13  &  6.48   & 1.25  &        6.20
       &  0.95     & 0.89        \\
AS4 & $-$219   & $-$52    & 139      &   4.64  &  7.11   & 1.14  &        7.14
       &  0.88     & 0.72        \\
AS5 & $-$223   & $-$69    & 137      &   4.64  &  6.90   & 1.17  &        9.82
       &  0.96     & 0.72        \\
LP1 & $-$152   & ~~~~8    & 177      &   7.14  &  5.64   & 0.83  &       14.22
       &  2.21     & 0.99        \\
LP2 & $-$221   & $-$64    & 110      &   5.15  &  5.10   & 0.74  &       15.49
       &  2.68     & 0.71        \\
LP3 & $-$230   & $-$72    & 108      &   5.20  &  5.48   & 0.86  &       13.09
       &  3.45     & 0.87        \\
MP1 & $-$ 99   & ~~~41    & 202      &   6.32  &  5.52   & 0.78  &       16.22
       &  2.32     & 0.84        \\
MP2 & $-$ 96   & ~~~35    & 205      &   5.10  &  5.79   & 0.83  &       17.02
       &  2.60     & 0.74        \\
MP3 & $-$161   & $-$17    & 130      &   5.49  &  5.43   & 1.03  &       12.69
       &  1.59     & 1.00        \\
MP4 & $-$191   & $-$34    & 132      &   4.64  &  4.91   & 1.04  &       15.43
       &  1.43     & 0.89        \\
RO1 & $-$209   & $-$69    & 110      &   4.55  &  6.68   & 0.79  &       18.50
       &   -       & 0.64        \\
RO2 & $-$193   & $-$46    & 130      &   3.65  &  5.27   & 0.95  &       12.13
       &   -       & 0.60        \\
RO3 & $-$215   & $-$88    & ~83      &   5.55  &  5.52   & 0.71  &       10.42
       &   -       & 0.54        \\
RO4 & $-$184   & $-$76    & 105      &   5.57  &  5.15   & 0.77  &       11.69
       &   -       & 0.68        \\
\hline
\end{tabular}
\end{table*}

In Fig. \ref{specplot} we show our normalized H$\alpha$ record of
\lsi\ ordered sequentially with radio phase and drawn on the same
scale with an arbitrary offset.
For days with multiple
measurements, only the best profile is shown.
The continuum underlying
H$\alpha$ was rectified to unity employing a spline fitting.  This
normalization procedure is somewhat arbitrary for Rozhen Observatory
spectra, since the spectral range covered is very small, and since most
of the H$\alpha$ wings is probably lost in noise.

The H$\alpha$ line profile of \lsi\ shows broad wings as well as a
double peaked core. The Full Width Zero Intensity (FWZI) of H$\alpha$
measured on the Mt. Palomar spectra is $\approx 3100$ km s$^{-1}$.
In addition, in the Asiago spectra,
the red hump is nearly flat topped and shows a broad shoulder
to the red.

Line parameters were measured on each normalized spectrum and are
reported in Table \ref{specpar}.  First column gives the spectrum
identification.  Second, third and fourth columns list the heliocentric
radial velocity of the blue peak, central dip, and red peak,
respectively.  Fifth and sixth columns are the FWHM of the blue and red
hump, corrected for instrumental profile, while seventh column provides
the ratio between the blue and red peak intensity above continuum.
Eighth column gives the total H$\alpha$ EW and the ninth column lists the
EW of the wings.  Finally, the tenth column contains the EW ratio of the
red and blue humps.

The radial velocity, the FWHM and the peak height of the B and R humps
were measured employing a gaussian fitting.  In the high resolution
spectra, the fitting was done after rebinning to a dispersion of
$\sim$1~\AA/pixel.
We estimated the contribution of
the wings by subtracting from the H$\alpha$ profile a model of the core
composed of two gaussians.

\section{Spectroscopic analysis} \label{specan}

\subsection{Radial velocities}

\begin{figure}
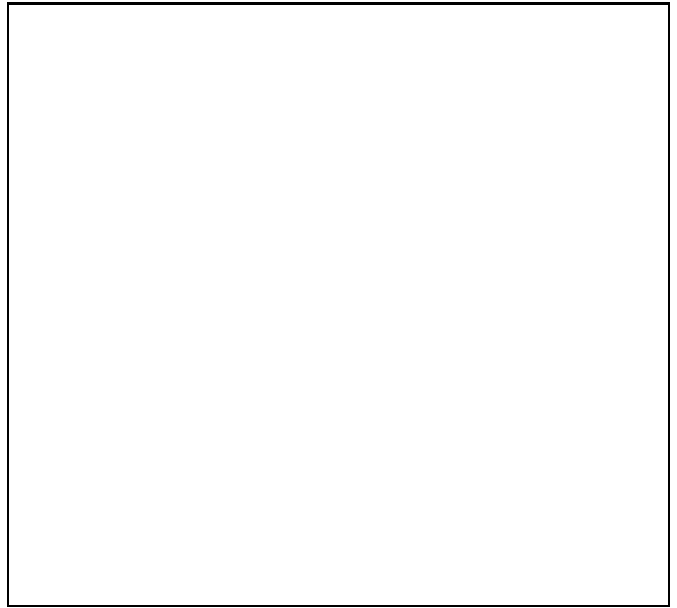

\picplace{8.0cm}
\caption{\label{vrad}  Radial velocity difference,
between the R and B peaks, and
heliocentric radial velocity of the central dip. Both are in \kms and
horizontal axis is labeled in radio phase.
The size of the dots is proportional to the dispersion,
i.e., the larger the dot, the higher the reliability of the measurement.}
\end{figure}

In Fig.  \ref{vrad}, we show the radial velocity difference between the
R and B peaks as well as the central dip radial velocity, both as a function
of radio phase.
The velocity difference between the H$\alpha$ peaks
reaches a maximum close to the radio outburst, during an interval
of two tenths of radio phase. On the other hand,
Gregory et al.  (1979) noted already that the Balmer
central dip varied within the range $-$30/$-$80 km s$^{-1}$ over the
period February-March 1978.  Our data confirm this variation of radial
velocity.  We also find that the central dip has $v_r > 0$ \kms,
but only in two low resolution spectra.
A global shift of the H$\alpha$
line is probably $\simlt$ 60 \kms, if we neglect the three "outsider"
points that come from low resolution, low significance observations.  A
weighted average of the radial velocities over the inverse square of
dispersion yields $v_r({\rm B}) \approx -208$ \kms, $v_r({\rm dip}) \approx
-60$ \kms.
The variation in the peak separation may be
due to an increase in $v_r$ of the red peak and a decrease in the
$v_r$ of the blue peak.

\subsection{Line width and {\rm B/R} variability}

\begin{figure}
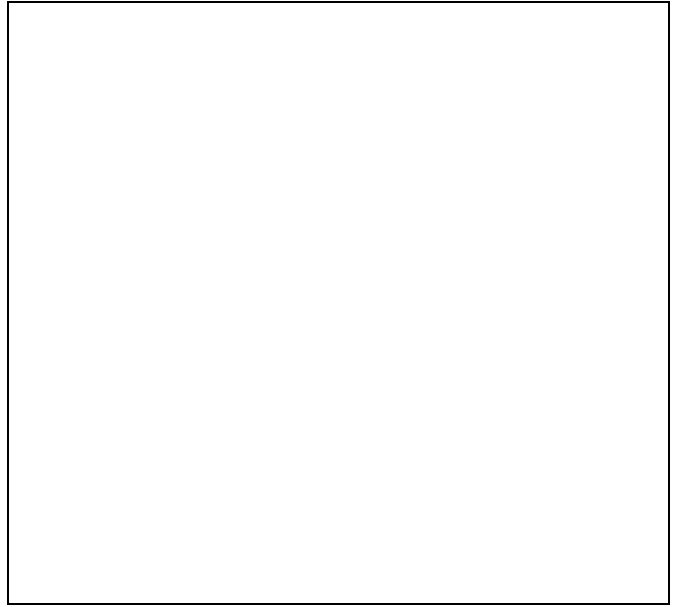

\picplace{8.00cm}
\caption{\label{fw}
FWHM(R) and FWHM(B) as a function of radio phase. Black dots represent
the FWHM of the fitting gaussians and white dots correspond to the
hump half maximum width.
The size of the dots is proportional to the dispersion.}
\end{figure}

\begin{figure}
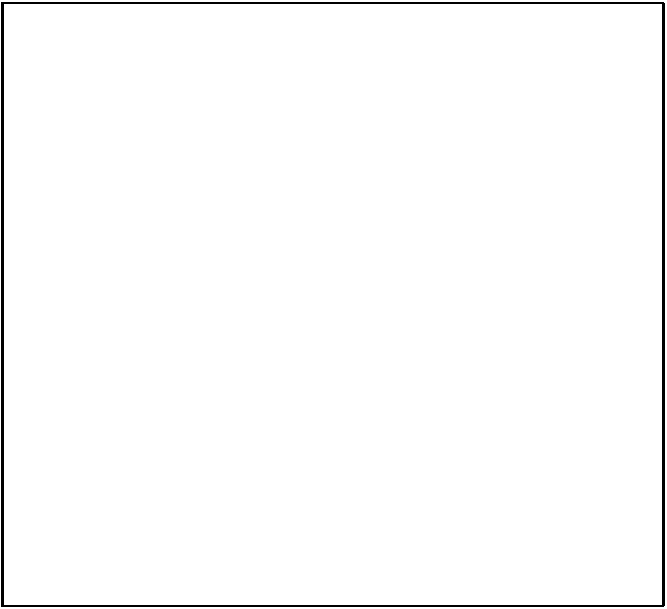

\picplace{8.00cm}
\caption{\label{brew}
B/R peak ratio and EW(H$\alpha$) in \AA~as a function of radio phase.}
\end{figure}

In Fig.  \ref{fw} we represent the FWHM(R) and FWHM(B) as a function of
radio phase, while Fig.  \ref{brew} illustrates the B/R peak ratio and
EW(H$\alpha$) as a function of radio phase also.

The H$\alpha$ profile of \lsi\ is not peculiar with respect to
those observed in other Be stars (we can, for instance, compare \lsi\ to
the Be stars studied by Slettebak et al. (1992)).  The
EW(H$\alpha$) of \lsi\ is $\simlt$ 20 \AA. This value sets \lsi\
among the H$\alpha$-weak Be stars.  However, the value of the
H$\alpha$ FWZI is, to the best of our knowledge, among the largest
values observed in Be stars.

Gregory et al.  (1979) described \lsi\ as having Balmer line emission of
variable profile intensity and decrement.  Later, HC81 showed that the
radial velocity of the central dip, the emission intensity, and the peak
ratio were all phase related quantities.  The H$\alpha$ line profile
did not change in its main features over 14 years.  The stability of the
H$\alpha$ profile is undoubtly remarkable, since Be stars sometimes
exhibit strong H$\alpha$ profile changes over timescales of several
months.  Our data confirm that the H$\alpha$ EW changes strongly in
correspondence of the radio outburst.  The minimum value of EW (6 \AA)
and the maximum value of peak ratio B/R are observed between radio phase
0.7-0.8 (see Fig. \ref{brew}).

As can be seen in Fig.  \ref{fw}, the FWHM(R) increases from a
value of $\sim$6 \AA~($\sim$250 \kms), at radio phase $\sim$0.5, to a
value of $\sim$8 \AA~($\sim$350 \kms) at radio phase $\sim$0.7~.  So,
the red hump seems to become substantially broader near the time of
radio maximum.  On the contrary, the FWHM(B) appears to decrease
slightly at the same time.
In the Asiago spectra
(obtained close to the radio maximum), the FWHM of the red hump is
visibly larger by $\approx$ 100 \kms than that of the blue one.
The exact values of the line width depends
somewhat on the method employed for the measurement.  To ascertain that
this effect is real we also measured directly the half width of the two
humps.  Although the numbers are somewhat different, the same effect is
evident in Fig. \ref{fw}.  Also, we have computed the flux ratio of the blue
and
red hump, which happens to be $\simlt 1$ at all epochs of observation.
This implies that the change in peak intensity ratio is mainly due to a
variation in width of the red hump.

The flux ratio between line core and line wings is relatively constant
in all spectra ($\sim 0.15$) but LP3.  The H$\alpha$ profile in
LP3 displays a prominent blue wing, with EW(wings)/EW(core)
$\sim 0.26$, and asymmetry index of the line wings AI$\sim$ $-$0.37,
where AI is defined as [EW(Red Wing) --EW(Blue Wing)]/EW(Both Wings).
In the other low resolution
observations, the H$\alpha$ line wings are symmetric within the
uncertainties.

\subsection{Spectroscopic discussion}  \label{specdis}

The most widely accepted explanation of the H$\alpha$ line profiles
in Be stars involves a circumstellar disk-like envelope that produces
the double-peaked line core.  Electron scattered H$\alpha$ photons
are expected to produce extended line wings.
The FWZI of \lsi\ H$\alpha$ is $\approx$ 3100 \kms\ $\gg 2v\sin{i} \sim
780$-$960$ \kms (HC81).
The excess in the line
wings is especially evident if we model the core as a sum of two
gaussians (FWZI(core) $\approx$ 1000 \kms).
If we assume that the line core is emitted in a
disk, and that the velocity field in the disk is keplerian, we can
estimate the ratio between the inner and outer radius of the disk.  We
obtain $R_{out}/R_{in} \approx 9.2$, for FWZI(core) $\approx 1000$ \kms.
The electron scattering optical depth $\tau_{es}$ for a disk-like
geometry can also be computed.  The density was assumed to depend upon
$r$ as $n_e = n_{e,0} (r/R_{in})^{-\alpha}$, and to fade exponentially
above and below the symmetry plane of the disk.  For $n_{e,0} = 10^{12}$
cm$^{-3}$, and for $\alpha=2$, we find that $\tau_{es} \simgt 0.3$, and,
if the density n$_{e,0}$ is $\simgt 10^{12}$ cm$^{-3}$, $\tau_{es} \sim
1$.  If $\tau_{es}$ is so large, and if the temperature of the gas is
$T_e \simgt 10^4$ K, as likely, extended line wings of FWZI $\sim$ 3000
\kms can be produced (Poeckert \&\ Marlborough, 1979).

The circumstellar disk around the B star should have $R_{out} \sim 5\times
10^{12}$ cm, a value similar to the length of  the semimajor axis of the
orbit estimated by HC81. If the eccentricity is $\approx 0.75$, the
secondary star should cross the circumstellar disk and, if the
circumstellar disk itself is nearly coplanar to the plane of the orbit,
even sweep across it while close to periastron.  The resultant accretion
could produce the radio outburst. A similar model has been proposed for the
Be star/X-ray binary systems A0538$-$66 and V0332$+$52 (e.g., Slettebak,
1988). The increase in width of the red hump could be due to a
non-axisymmetric perturbation in the circumstellar disk, occurring close to
the outer edge of the disk. For instance, the secondary may be crossing the
circumstellar shell at that time, close to its outer radius (it is
interesting to note that the blue hump is probably wider than the red one
at radio phase $\approx 0.4$). However, the orbital solution of HC81 suggests
that the broadening is occurring when the star is close to apoastron,
where the secondary is probably not in contact with the circumstellar disk
of the Be star, if the radius of the circumstellar disk is $R_{out} \approx
5 \times 10^{12}$ cm.

Alternatively, most of the H$\alpha$ emission could be associated with
the compact secondary.  An accretion disk around a $\sim$1 M$_\odot$
compact object would have $R_{out} \sim 1 \times 10^{12}$ cm,
without considering broad wings produced by electron scattering.
The strong decrease in the H$\alpha$ EW at radio maximum can be
explained either in terms of obscuration of the disk by the Be star
(Mendelson \& Mazeh, 1989), or in terms of reduced emissivity in the disk.

The JHK light curves point toward an eclipsing binary.  Hence,
the plane of the orbit should be close to the line of sight.  Since the
secondary has probably a mass of only 1/6 or less the mass of the
primary, we expect also a large radial velocity oscillation in the peak
positions ($\sim$ 300 \kms).  Our radial velocity data are not
consistent with this, nor are the data obtained by the
previous investigators.

Less clear is the interpretation of
the variation of the wings; we think that a set of homogeneous observations
of sufficiently high S/N and resolution are needed to finally
reject the possibility that the line wings might be emitted by the
accretion disk of the secondary.

Even if the accretion disk around the compact star does not emit the
bulk H$\alpha$ luminosity, the increase in FWHM(R) and in $\Delta v_r
= v_r({\rm R}) - v_r({\rm B})$ may be caused by an unresolved component whose
radial
velocity reaches a maximum in correspondence to the radio maximum.  We
may expect this if, for instance, a cloud of line emitting gas would be
ballistically ejected along the secondary disk axis.  This suggestion is
appealing, since the radial velocity of this unseen component should be
$\simgt$ 200 \kms at radio maximum, consistent with the velocity of the
bipolar ejecta of radio plasma partially resolved with VLBI techniques
by Massi et al.  (1993).  A strong analogy could be envisaged with the
model proposed by Martin \& Rees (1979) for SS433.  Present data on
radial velocity and FWHM variation can be explained by a combination of
global line displacement due to the orbital motion of the Be star, and
of the $v_r$ variation of this unresolved component.  The red shoulder
often present in the H$\alpha$ profile could be a related, higher
velocity, feature.

\section{Conclusions} \label{conclu}

Our Johnson photometric monitoring of \lsi\ has shown that this object
presents a $\sim$26 d periodic modulation in the V band.
The shape and amplitude of this modulation are similar to those
found by Mendelson \& Mazeh (1989).  In addition, the J, H
and K observations reported in this paper have revealed, for the first
time, new evidence of infrared variability with similar trends as seen
in the optical, but with higher amplitude (0\rmag 2).  We have
established also that the merged JHK light curve exhibits a modulation
with period similar to the radio period.  A possible interpretation of
this periodicity could involve the eclipse and attenuation of the
secondary star emission by the Be primary and its envelope.

It is unclear whether an accretion disk around a compact,
degenerate companion may contribute to the H$\alpha$ emission,
or whether the variations observed are due to perturbations produced by the
companion on the circumstellar disk of the Be primary.
This problem persists also because Be stars, as
a class, are far from being well understood.  \lsi\ clearly deserves more
observations from ground and space.  Monitoring of the H$\alpha$
profile at high and intermediate resolution will help to solve the main
ambiguities left by the present investigation.  It is also desiderable
to obtain a new spectroscopic orbital solution.

\acknowledgements{We thank all observers who also collaborated in the
observations, especially Mauro D' Onofrio \& Gabriele Cremonese for
the spectra obtained at the Asiago Observatory, and for help during part
of the spectroscopic data reduction.  We also thank R. Canal for
valuable comments as well as F. Comer\'on and M. Fern\'andez
for participating in some
photometric observations.  JMP, JM, FF, CJ and JT acknowledge
partial support by CICYT (ESP93-1020-E) and DGICYT (PB91-0857).
Extragalactic Astronomy at the University of Alabama is
supported under EPSCOR Grant R11--8996152.  The TCS is operated on the
island of Tenerife by the Instituto de Astrof\'{\i}sica de Canarias (IAC), in
the
Spanish Observatorio del Teide.  The 1.5~m telescope at Palomar Mt. is
jointly owned by California Institute of Technology and the Carnegie
Institute of Washington. The CAHA 1.23 m telescope is operated by Max Planck
Institute f\"ur Astronomie.
The INT and JKT are operated on the island of La Palma
by the Royal Greenwich Observatory in the Spanish ORM of the IAC.
We also thank the staff of the OAN 1.5 m telescope.
Much of the data were analysed using the Southampton University
Starlink node which is founded by the SERC. CE acknowledges the support
of an SERC Studentship. We are also grateful for the support of the ING
Service Programme that provided some of the data for this work.
}

\end{document}